\magnification=1200
\def\qed{\unskip\kern 6pt\penalty 500\raise -2pt\hbox
{\vrule\vbox to 10pt{\hrule width 4pt\vfill\hrule}\vrule}}
\centerline{A THEORY OF HYDRODYNAMIC TURBULENCE}
\centerline{BASED ON NON-EQUILIBRIUM STATISTICAL MECHANICS.}
\bigskip\bigskip\bigskip\bigskip
\centerline{by David Ruelle\footnote{$\dagger$}{Math. Dept., Rutgers University, and 
IHES, 91440 Bures sur Yvette, France. email: ruelle@ihes.fr}.}
\bigskip\bigskip\bigskip\bigskip\noindent
	{\leftskip=2cm\rightskip=2cm\sl Abstract. In earlier papers, we have studied the turbulent flow exponents $\zeta_p$, where $\langle|\Delta{\bf v}|^p\rangle\sim\ell^{\zeta_p}$ and $\Delta{\bf v}$ is the contribution to the fluid velocity at small scale $\ell$.  Using ideas of non-equilibrium statistical mechanics we  have found
$$	\zeta_p={p\over3}-{1\over\ln\kappa}\ln\Gamma({p\over3}+1)   $$
where $1/\ln\kappa$ is experimentally $\approx0.32\pm0.01$.  The purpose of the present note is to propose a somewhat more physical derivation of the formula for $\zeta_p$.  We also present an estimate $\approx100$ for the Reynolds number at the onset of turbulence.\par}
\vfill\eject
\null\bigskip\bigskip
	In previous papers ([6],[7],[8]) we have used ideas of non-equilibrium statistical mechanics to study hydrodynamic turbulence.  In particular we have analyzed the exponents $\zeta_p$, where $\langle|\Delta{\bf v}|^p\rangle$ $\sim\ell^{\zeta_p}$ and $\Delta{\bf v}$ is the contribution to the fluid velocity at small scale $\ell$.  These exponents were discussed earlier in [2],[3].  Here we find these exponents $\zeta_p$ to satisfy
$$	\zeta_p={p\over3}-{1\over\ln\kappa}\ln\Gamma({p\over3}+1)\eqno{(1)}   $$
where $1/\ln\kappa$ is experimentally $\approx0.32\pm0.01$.  In this note we give a more physical presentation of our basic ideas (Section 1), then derive equation (1) (Section 2).  We also discuss briefly the distribution of radial velocity increments (Section 3).  Finally we give an estimate $\approx100$ for the Reynolds number at the onset of turbulence (Section 4).
\bigskip
	{\bf 1. Introduction and basic probability distribution.}
\medskip
	We consider a viscous incompressible fluid in 3 dimensions where energy is input at a large spatial scale $\ell_0$.  The energy then goes down to small spatial scales through a cascade of eddies of increasing order $n$ so that the fluid velocity field ${\bf v}$ is
$$	{\bf v}=\sum_{n\ge0}{\bf v}_n   $$
and the sum is cut off by viscous dissipation when the size of the $n$-th order eddies is small enough.
\medskip
	We assume that (starting from a ``zero-th order eddy'' of size $\ell_0$) an eddy of order $n-1$ contained in a ball $R_{(n-1)i}$ decomposes after time $T_{(n-1)i}$ into eddies of order $n$ contained in balls $R_{nj}\subset R_{(n-1)i}$.  The balls $R_{nj}$ of a given order form a partition of 3-space into roughly spherical polyhedra of linear size $\approx\ell_{nj}$.  [In earlier papers we took the $R_{nj}$ to be cubes with side $\ell_n=\ell_0\kappa^{-n}$].
\medskip
	We fix the choice of the balls $R_{nj}$ of various orders (we shall later integrate over this choice) and discuss the fluctuating distribution of the kinetic energies $E(R_{nj})$ of the eddies.  We assume that the dynamics of each eddy is universal, up to scaling of space and time, and independent of other eddies.  Let the lifetime of the eddy in $R_{nj}$ be $T_{nj}$.  When the eddy in $R_{(n-1)i}$ decomposes into daughter eddies in $R_{nj}$, conservation of (kinetic) energy $E$ yields
$$	\sum_j{E(R_{nj})\over T_{nj}}={E(R_{(n-1)i})\over T_{(n-1)i}}\eqno{(2)}   $$
Universality of the dynamics and the scaling properties of the inviscid evolution equations then give
$$	{{\bf v}_n\over\ell_{nj}}={T_{(n-1)i}\over T_{nj}}\cdot{{\bf v}_{n-1}\over\ell_{n-1}}   $$
for the initial velocity fields ${\bf v}_n$, ${\bf v}_{n-1}$ of the eddies in $R_{nj}$, $R_{(n-1)i}$.  We choose the definition of $\ell_{nj}$, $\ell_{(n-1)i}$ so that this relation is exact, therefore (2) gives
$$	\sum_j\int_{R_{nj}}{|{\bf v}_n|^3\over\ell_{nj}}=\int_{R_{(n-1)i}}{|{\bf v}_{n-1}|^3\over\ell_{(n-1)i}}\eqno{(3)} $$
$\Big[$Instead of using the initial velocity fields ${\bf v}_n$, ${\bf v}_{n-1}$ we could also average the integrals in (3) over time in $[0,T_{nj}]$, $[0,T_{(n-1)i}]$ $\Big]$.
\medskip
	Note that
$$	\sum_j\int_{R_{nj}}|{\bf v}_n|^2=\int_{R_{(n-1)i}}|{\bf v}_{n-1}|^2\eqno{(4)}   $$
If ${\bf v}_{(n-1)}$ and ${\bf v}_n$ are nearly constant in $R_{(n-1)i}$ and $R_{nj}$ respectively they must take quite different values because $\ell_{nj}<\ell_{(n-1)i}$ in (3).  Therefore (3) and (4) imply big fluctuations of ${\bf v}_n$ between the different $R_{nj}$ (this is intermittency).
\medskip
	For simplicity we shall assume that the fluctuations in size of the $R_{nj}$ for different $j$ are not important (even if there are big fluctuations of ${\bf v}_n$ between the $R_{nj}$).  In particular we shall let $\ell_{(n-1)i}/\ell_{nj}=\kappa$ independently of $i,j$.  Therefore (3) gives
$$	\kappa\sum_j\int_{R_{nj}}|{\bf v}_n|^3=\int_{R_{(n-1)i}}|{\bf v}_{n-1}|^3\eqno{(5)}   $$
We make also the essential assumption that the distribution of the ${\bf v}_n$ between the different $R_{nj}$ maximizes the entropy so that the integrals of $|{\bf v}|^3$ over the different $R_{nj}$ satisfy a microcanonical distribution with respect to an ``energy'' $|{\bf v}_n|^3$.  We can replace this by a canonical distribution
$$	\sim\exp[-\beta|{\bf v}_n|^3]\,d^3{\bf v}_n   $$
Integrating over angular variables this is
$$ \sim\exp[-\beta|{\bf v}_n|^3]\,|{\bf v}|^2\,d|{\bf v}_n|={1\over3}\exp[-\beta|{\bf v}_n|^3]\,d{|\bf v}_n|^3\eqno{(6)} $$
\indent
	Defining $V_n=|{\bf v}_n|^3$, which we take for simplicity to be constant in each $R_{nj}$, we find that $V_n$ has a distribution
$$	\beta\exp[-\beta V_n]\,dV_n\eqno{(7)}   $$
To obtain (6) and (7) it is crucial that the vector ${\bf v}$ is 3-dimensional.  Using (5) we find that the average value $\beta^{-1}$ of $V_n$ is $V_{n-1}/\kappa$ so that $V_n$ is distributed on $(0,\infty)$ according to
$$	{\kappa\over V_{n-1}}\exp\Big[-{\kappa V_n\over V_{n-1}}\Big]\,dV_n   $$
\Big[In the above argument we imagine $|{\bf v}_n|$ to be initially constant in each $R_{nj}$, but we could also take $V_n$ to be some average of $|{\bf v}_n|^3$\Big].
\medskip
	Starting from a given value of $V_0$ the distribution of $V_n$ is thus given by
$$	{\kappa dV_1\over V_0}e^{-\kappa V_1/V_0}\cdots
	{\kappa dV_n\over V_{n-1}}e^{-\kappa V_n/V_{n-1}}\eqno{(8)}   $$
Integrating over the choice of the $R_{nj}$ with fixed $\ell_{(n-1)i}/\ell_{nj}=\kappa$ does not change the formula (8).
\bigskip
	{\bf 2. Calculating ${\zeta}_p$.}
\medskip
	We compute now the mean value of $|{\bf v}_n|^p=V_n^{p/3}$.  We note that
$$	{\kappa\over V_{n-1}}\int\exp\Big[-{\kappa V_n\over V_{n-1}}\Big].V_n^{p/3}\,dV_n
	=\Big({V_{n-1}\over\kappa}\Big)^{p/3}\int\exp[-w].w^{p/3}\,dw   $$
$$	=\kappa^{-p/3}V_{n-1}^{p/3}\Gamma\Big({p\over3}+1\Big)   $$
Therefore, using induction, we find
$$	\langle V_n^{p/3}\rangle={\kappa\over V_0}\int\exp\Big[-{\kappa V_1\over V_0}\Big]\,dV_1\cdots
	{\kappa\over V_{n-1}}\int\exp\Big[-{\kappa V_n\over V_{n-1}}\Big].V_n^{p/3}\,dV_n   $$
$$	=\kappa^{-np/3}V_0^{p/3}\Gamma\Big({p\over3}+1\Big)^n
	=V_0^{p/3}\Big({\ell_n\over\ell_0}\Big)^{p/3}\Gamma\Big({p\over3}+1\Big)^n   $$
\indent
	We have ignored the fluctuations of $\ell_n$, so that we could write $\ell_n/\ell_0=\kappa^{-n}$ and
$$	\ln\langle|{\bf v}_n|^p\rangle=\ln\langle V_n^{p/3}\rangle
=	\ln V_0^{p/3}+{p\over3}\ln\Big({\ell_n\over\ell_0}\Big)
	-{\ln(\ell_n/\ell_0)\over\ln\kappa}\ln\Gamma\Big({p\over3}+1\Big)   $$
$$=\ln V_0^{p/3}+\ln\Big({\ell_n\over\ell_0}\Big)
	.\Big[{p\over3}-{1\over\ln\kappa}\ln\Gamma\Big({p\over3}+1\Big)\Big]
	=\ln\Big[V_0^{p/3}\Big({\ell_n\over\ell_0}\Big)^{\zeta_p}\Big]   $$
where
$$	\zeta_p={p\over3}-{1\over\ln\kappa}\ln\Gamma\Big({p\over3}+1\Big)   $$
or
$$	\langle|{\bf v}_n|^p\rangle=V_0^{p/3}\Big({\ell_n\over\ell_0}\Big)^{\zeta_p}\sim\ell_n^{\zeta_p}   $$
as proposed in [6],[7],[8].\footnote{(*)}{The formula (1) for $\zeta_n$ yields results for $n$ up to 10 close to those obtained by Victor Yakhot by an approach which is apparently quite different (see [10]).  V.Y. also points out that for n large ($n>50$), (1) violates the H\"older inequality, and can no longer be trusted.}
\medskip
	[It would be possible to take into account the fluctuations of $\ell_n$ and write a formula for $\ln\langle|{\bf v}_n|^p\rangle$ integrating over those fluctuations].
\bigskip
{\bf 3. Estimating the distribution $F(u)$ of the radial velocity increment $u$.}
\medskip
	In this Section we follow [7],[8].  We write $u=\Delta_rv=\sum_mu_m$ and if $r\approx\ell_n$ we have $u\approx u_n=\hbox{radial component of }{\bf v}_n$.  Therefore, given $V_0$, a rough estimate of $F(u)$ is given by
$$	F(u)=\Big(\prod_{m=1}^n\int_0^\infty{\kappa\,dV_m\over V_{m-1}}e^{-\kappa V_m/V_{m-1}}\Big)
	{1\over2V_n^{1/3}}\chi_{[-V_n^{1/3},V_n^{1/3}]}(u)   $$
$$	={1\over2}\Big({\kappa^n\over V_0}\Big)^{1/3}\int\cdots\int_{w_1\cdots w_n>(\kappa^n/V_0)|u|^3}
	\prod_{m=1}^n{dw_m\,e^{-w_m}\over w_m^{1/3}}   $$
Instead of $F(u)\,du$ we may consider the distribution $G_n(y)\,dy$ of $y=(\kappa^n/V_0)^{1/3}|u|$ so that
$$	G_n(y)=\int\cdots\int_{w_1\cdots w_n>y^3}\prod_{m=1}^n{dw_m\,e^{-w_m}\over w_m^{1/3}}   $$
and one finds
$$	e^tG_n(e^t)=(\phi^{*(n-1)}*\psi)(t)\eqno{(9)}   $$
with
$$	\phi(t)=3\exp(3t-e^{3t})\qquad,\qquad\psi(t)=e^t\int_t^\infty e^{-s}\phi(s)\,ds\eqno{(10)}   $$
One can show that $G_n(y)$ is a decreasing function of $y$.  Also, $F(u)$ gives a reasonable fit of the numerical data [9] for small $u$.
\medskip
	The formula (9) suggests a lognormal distribution with respect to $u$, in agreement with Kolmogorov [5] and Obukhov.  However (10) shows that $\phi,\psi$ tend only exponentially to 0 at $-\infty$, which is not sufficient to obtain a lognormal distribution and the prediction of $\zeta_n$ made by the lognormal theory.  It is therefore satisfactory that (1) gives a better fit to the experimental data of [1].\footnote{(*)}{See the paper [4] by Giovanni Gallavotti and Pedro Garrido for a discussion of the relation of (9),(10) to Kolmogorov-Obukhov.}
\bigskip
{\bf 4. The onset of turbulence.}
\medskip
	We may estimate the Reynolds number ${\cal R}e=|\bf v_0|\ell_0/\nu$ for the onset of turbulence by taking
$$	1\approx\Big\langle{\nu\over|{\bf v}_1|\ell_1}\Big\rangle
	=\Big\langle{\nu\over V_1^{1/3}\kappa_{-1}\ell_0}\Big\rangle
	={\cal R}e^{-1}\Big\langle\kappa^{4/3}\Big({V_0\over\kappa V_1}\Big)^{1/3}\Big\rangle   $$
This relation to dissipation is dictated by dimensional arguments and corresponds to
$$	{\cal R}e\approx
\kappa^{4/3}\int_0^\infty\Big({\kappa V_1\over V_0}\Big)^{-1/3}{\kappa\,dV_1\over V_0}\,e^{-\kappa V_1/V_0}
	=\kappa^{4/3}\int_0^\infty\alpha^{-1/3}\,d\alpha\,e^{-\alpha}
	=\kappa^{4/3}\Gamma\Big({2\over3}\Big)   $$
Taking $1/\ln\kappa=.32$ hence $\kappa^{4/3}=64.5$, with $\Gamma(2/3)\approx1.354$ gives ${\cal R}e\approx87$ which agrees with ${\cal R}e\approx100$ as found in [9].
\bigskip
	{\bf Acknowledgements.}
\medskip
	I am indebted to Giovanni Gallavotti, Pedro Garrido, and Victor Yakhot for useful discussions about the present paper.
\bigskip
	{\bf References.}

[1] F. Anselmet, Y. Gagne, E.J. Hopfinger, and R.A. Antonia.  ``High-order velocity structure functions in turbulent shear flows.''  J. Fluid Mech. {\bf 140},63-89(1984).

[2] R. Benzi, G. Paladin, G. Parisi, and A. Vulpiani.  ``On the multifractal nature of fully developed turbulence and chaotic systems.''J. Phys. A {\bf 17},3521-3531(1984).

[3] U. Frisch and G. Parisi ``On the singularity structure of fully developed turbulence'' in {\it Turbulence and Predictability in Geophysical Fluid Dynamics} (ed. M. Ghil, R. Benzi, and G. Parisi), pp. 84-88.
North-Holland, 1985.

[4] G. Gallavotti and G. Garrido.  ``Non-equilibrium statistical mechanics of turbulence: comments on Ruelle's intermittency theory'' Chapter 4 (pp 59-70) in Chr. Skiadas (editor)  {\it The foundations of chaos revisited: from Poincar\'e to recent advancements.}  Springer, Heidelberg, 2016.	

[5] A.N. Kolmogorov.  ``A refinement of previous hypotheses concerning the local structure of turbulence in a viscous incompressible fluid at high Reynolds number.''  {\it J. Fluid Mech.} {\bf 13},82-85(1962).

[6] D. Ruelle.  ``Hydrodynamic turbulence as a problem in nonequilibrium statistical mechanics.''  PNAS {\bf 109},20344-20346(2012).

[7] D. Ruelle.  ``Non-equilibrium statistical mechanics of turbulence.''  J. Statist. Phys. {\bf 157},205-218(2014).

[8] D. Ruelle.  ``Hydrodynamic turbulence as a nonstandard transport phenomenon.''  Chapter 3 (pp 49-57) in
Chr. Skiadas (editor)  {\it The foundations of chaos revisited: from Poincar\'e to recent advancements.}  Springer, Heidelberg, 2016.	

[9] J. Schumacher, J. Scheel, D. Krasnov, D. Donzis, K. Sreenivasan, and V. Yakhot.  ``Small-scale universality in turbulence.''  PNAS {\bf 111},10961-10965(2014).

[10] V. Yakhot and D. Donzis. ``Emergence of multiscaling in a random-force stirred fluid.''  arXiv:1702.08468.
\end